# A Highly Correlated Topological Bubble Phase of Composite Fermions

V. Shingla[1]†, Haoyun Huang[1]†, A. Kumar[2], L.N. Pfeiffer[3], K.W. West[3], K.W. Baldwin[3], and G.A. Csáthy[1]*

[1]Department of Physics and Astronomy, Purdue University; West Lafayette, Indiana 47907, USA

[2]Department of Physics, Monmouth College; Monmouth, Illinois 61462, USA

[3]Department of Electrical Engineering, Princeton University; Princeton, New Jersey 08544, USA

†These authors contributed equally    *Corresponding author. Email: gcsathy@purdue.edu

**Strong interactions and topology drive a wide variety of correlated ground states. Some of the most interesting of these ground states, such as fractional quantum Hall states and fractional Chern insulators, have fractionally charged quasiparticles. Correlations in these phases are captured by the binding of electrons and vortices into emergent particles called composite fermions. Composite fermion quasiparticles are randomly localized at high levels of disorder and may exhibit charge order when there is not too much disorder in the system. However, more complex correlations were predicted when composite fermion quasiparticles cluster into a bubble, then these bubbles order on a lattice. Such a highly correlated ground state was termed the bubble phase of composite fermions. Here we report the observation of this bubble phase of composite fermions, evidenced by the reentrance of the fractional quantum Hall effect. We associate this reentrance with a bubble phase with two composite fermion quasiparticles per bubble. Our results demonstrate the existence of a new class of strongly correlated topological phases driven by clustering and charge ordering of emergent quasiparticles.**

Landau's symmetry breaking paradigm provides a framework to classify phases described by local order parameters. Topological phases do not fit into this classification and are described instead by topological invariants. Topological phases are characterized by the formation of edge states and of an insulating bulk and, in the vast majority of cases, symmetry breaking does not play any role. Indeed, the bulk of most ordinary topological phases, such as the integer quantum Hall state forming in the two-dimensional electron gas (2DEG)[1], is an Anderson insulator. As the Landau level filling factor moves away from an integer value, bulk quasiparticles (QPs) are generated, which are randomly localized. Local scanning probes provided evidence for such randomly localized bulk QPs[2]. A representation of integer quantum Hall states with a finite QP density is shown in Fig.1*a*.

In addition to ordinary topological phases characterized solely by topological invariants, there is a larger class of phases for which both topological and Landau-type of orders need to be invoked. Such phases exhibit topologically protected edge states, while QPs in their bulk break various spatial symmetries. An example of such a phase is the Wigner solid (WS) forming in the flanks of integer quantum Hall states[3], a phase related to the WS forming in the extreme quantum limit[4]. In the limit of no disorder, quasiparticles in the bulk of this phase are thought to order on a triangular lattice, while edge states maintain integer quantization of the Hall resistance. So far conditions of low enough disorder for the formation of these types of WSs were met in 2DEGs confined to GaAs[3-9] and in graphene[10]. Even though microscopic observation of charge order is still lacking, measurements of the pinning mode[3], NMR charge topography[5], the phonon mode[6], and localization[7-10] provide evidence of charge ordering in the bulk.

The complexity of charge order is known to increase in high Landau levels, where more intricate broken symmetry topological phases are possible. Electronic bubble phases (EBPs) share the triangular lattice structure of the WS but acquire an internal degree of freedom: each lattice node consists of clusters or bubbles of electron-like QPs[11,12]. A representation of EBPs in the limit of no disorder with one and two QPs per bubble is shown in Fig.1$b$ and Fig.1$c$, respectively. Clearly, the one-QP EBP is identical to the WS. EBPs were predicted based on Hartree-Fock calculations performed in high Landau levels[11,12] and were later found to proliferate in 2DEGs hosted in GaAs[13-18]. More recently, EBPs were also found in graphene[19]. Clustering of electrons into bubbles is energetically favorable because of the existence of nodes in the overlapping electronic wavefunctions[11,12,18].

Phases discussed so far support electron-like QPs. Some of the most intriguing topological phases, such as fractional quantum Hall states (FQHSs)[20-22] and fractional Chern insulators[23-25] support fractionally charged QPs that emerge from strong correlations. Such correlations were first captured by Laughlin's wavefunction[26] and later found a natural and intuitive description within the theory of composite fermions (CFs)[27]. CFs are the emergent particles of this theory that form through binding of an even number of quantized vortices to electrons. CFs experience an effective magnetic field that much reduced when compared to the externally applied field. In this description, the density of states of CFs consists of equally spaced energy levels called Λ-levels, and FQHSs arise as CFs fill an integer number of Λ-levels[27]. When a Λ-level is not fully filled, CFQP are generated[27]. In the bulk of these FQHSs, the CFQPs are randomly localized[28], as depicted in Fig.1 $\tilde{a}$.

When CFQPs interact in a low disorder environment, the bulk may acquire charge order and therefore broken symmetry topological phases with a highly correlated nature may form. Such interactions are especially important in flatband systems, such as the 2DEG in magnetic field. Indeed, due to their interactions, CFQPs were predicted to order into a Wigner solid of CFs (WSCF)[29]. A representation of the WSCF is shown in Fig.1 $\tilde{b}$. Several numerical simulations found evidence for the formation of WSCFs[29-37]. However, because WSCFs and WSs exhibit a similar insulating behavior, these phases cannot readily be distinguished by commonly employed experimental probes. Resonances detected in the microwave frequency domain in the vicinity of $\nu = 1/3$ were interpreted as due to a WSCFs[38].

It is important to appreciate that the WSCF and the WS are not identical. Indeed, the WSCF has more intricate quantum mechanical correlations as compared to those of the WS. These correlations are depicted in Fig.1 $\tilde{b}$ as two vortices attached to each electron and are embodied by a distinctive Laughlin-Jastrow term present in the many-body trial wavefunction of the WSCF[29]. The vortex attachment procedure can therefore be understood as a recipe for generating topological phases with strong, higher order quantum mechanical correlations. Charge order with even more complex correlations was predicted when the vortex attachment procedure is applied to EBPs. The resulting phases are referred to as bubble phases of composite fermions (BPCFs)[39,40]. A rendering of BPCFs with one- and two-CFQPs per bubble is shown in Fig.1 $\tilde{b}$ and Fig.1 $\tilde{c}$, respectively.

**Observation of the Reentrance of the Fractional Quantum Hall State**

We focus on magnetotransport in the fractional quantum Hall regime in the lowest Landau level, in the region centered onto $v = 3/2$. Measurements are performed on a sample of density $n = 3.06 \times 10^{11} cm^{-2}$ and mobility $\mu = 32 \times 10^{6} cm^{2}/Vs$. Numerous FQHSs of this region form at filling factors of the form $v = 2 - i/(2i+1) = (3i+2)/(2i+1)$, where $i$ is an integer[41]. The FQHSs at $v = 5/3, 8/5, 11/7, 14/9, 17/11, 20/13$ seen in Fig.2 belong to this sequence, with $i = 1, 2, 3, 4, 5, 6$. These FQHSs are particle-hole symmetric counterparts of the ones forming at $v = i/(2i + 1)$ in the presence of the spin degrees of freedom. These FQHSs are identified by their vanishing longitudinal resistance $R_{xx} = 0$ and a Hall resistance quantized to the value $R_{xy} = (2i+1)h/(3i+2)e^{2}$. According to the CF theory, the above sequence of FQHSs forms when spinful two-flux CFs fill an integer number of $i = 1, 2, 3, 4, 5, 6$ of $\Lambda$-levels[27,41]. When the lowest $\Lambda$-level is completely filled, $i = 1$ and the ground state of the system is the $v = 5/3$ FQHS. Similarly, when two $\Lambda$-levels are completely filled, $i = 2$ and one obtains the FQHS at $v = 8/5$. Another series of FQHSs forms at $v = 2 - i/(2i-1)$, where $i = 2, 3, 4, 5, 6, 7$. Yet other FQHSs seen in Fig.2, such as the ones at $v = 9/7, 14/11$ and $12/7$, do not belong to these series. The local minimum in $R_{xx}$ near $B = 7.06$ T does not signal a FQHS; instead, it was recently associated with the WS[9].

Transport near $B = 7.76$ T exhibits a particularly interesting feature that breaks the typical pattern between two neighboring FQHSs. Indeed, neighboring FQHSs in the lowest Landau level are typically separated by a single peak in $R_{xx}$[20]. This is the case for data shown in Fig.2 for the transition between the $v = 8/5$ and the $v = 11/7$ FQHSs, the $v = 11/7$ and the $v = 14/9$ FQHSs, and for numerous other neighboring FQHSs. However, transport in the transition region between the $v = 5/3$ and the $v = 8/5$ FQHSs, i.e. near $B = 7.76$ T, is more complex: there are two $R_{xx}$ peaks that are separated by a vanishingly small $R_{xx}$. Furthermore, as seen in Fig.2, the Hall resistance $R_{xy}$ at $B = 7.76$ T at the lowest temperature is quantized to $3h/5e^{2}$. These details may be further examined in Fig.3. Altogether, we report four instances of similar complex transport; three of these can be found in the Supplement, where we demonstrate reproducibility after thermal cycling and an observation in a second sample. In the Supplement we also discuss anomalies in this filling factor region reported in prior work.

A complex transport behavior between two consecutive FQHSs can be due to either a spin transition or a novel ground state. However, the transport behavior observed near $B = 7.76$ T is inconsistent with a spin transition for three reasons. First, a quantized Hall resistance we observe at $B = 7.76$ T is not expected near a spin transition[41,42]. Second, the pattern of the longitudinal resistance measured near $B = 7.76$ T is different from that at a spin transition[41,42]. Third, a spin transition is not expected in the $\nu = 5/3$ FQHS[41], but it is known to occur in the $\nu = 8/5$ FQHS[42,43]. However, this transition is strongly dependent on the width of the confining quantum well[42]. The density $3.06 \times 10^{11} cm^{-2}$ of our 30 nm wide quantum well samples greatly exceed the critical density at this width[42], hence the $\nu = 8/5$ FQHS in our sample forms deeply in the fully spin polarized regime, far away from a spin transition. By ruling out a spin transition near $B = 7.76$ T, we ascertain that at this field there is a novel ground state forming.

For insight on the unusual transport pattern near $B = 7.76$ T, we recall the transport phenomenology of the reentrant integer quantum Hall states (RIQHSs). These states are satellite formations near integer quantum Hall states associated with electronic EBPs[13-19]. Both integer quantum Hall states and RIQHSs are characterized by a vanishing $R_{xx}$ and a quantized $R_{xy}$, but are separated from each other by a deviation from quantization. Transport behavior near $B = 7.76$ T is similar: both the $\nu = 5/3$ FQHS and the region near $B = 7.76$ T are characterized by a vanishing $R_{xx}$ and a quantized $R_{xy}$, and they are separated by a deviation from quantization developing near $B = 7.73$ T. However, in contrast to RIQHSs, the Hall resistance near $B = 7.76$ T is not quantized to an integer but rather to a fractional value $R_{xy} = 3h/5e^2$. Henceforth we will refer to the unusual transport developing near $B = 7.76$ T as the reentrant fractional quantum Hall state (RFQHS) and we associate it with a novel ground state of the system. From data shown in Fig.4 we infer that transport signatures for the RFQHS survive in our sample to temperatures as high as 60 mK.

**Candidate Ground States for the Reentrant Fractional Quantum Hall State**

To understand the nature of the RFQHS, we invoke the CF theory[27]. As already discussed, at $\nu = 5/3$ there is a FQHS. At this filling factor the two-flux CFs completely fill the lowest Λ-level. An increasing $B$ field in this region of filling factors results in a decrease of the effective magnetic field, which in turn leads to a decrease the degeneracy of this Λ-level. An increasing $B$ field past its value at $\nu = 5/3$ therefore generates CFQPs that reside in the second Λ-level. At relatively low densities, these CFQPs of the second Λ-level are Anderson-localized by the disorder present, contributing thus to the plateau of the $\nu = 5/3$ FQHS.

Near $B = 7.73$ T, $R_{xx}$ deviates from zero and $R_{xy}$ deviates from quantization. Such transport signatures are associated with delocalized CFQPs in the bulk. However, at an even larger magnetic field $B = 7.76$ T, there is a return to a nearly vanishing $R_{xx}$, a quantized $R_{xy}$ and thus to localization. However, localization near $B = 7.76$ T is inconsistent with an Anderson-type of localization. Instead, the observed reentrant behavior constitutes evidence for a correlated CFQP insulator in which localization is generated by pinning of a charge-ordered phase. A quantized $R_{xy} = 3h/5e^2$ of both the correlated CFQP insulator forming at $B = 7.76$ T and of the $\nu = 5/3$ FQHS suggests the same CFQPs are involved in the formation of both ground states. We thus identify the RFQHS with a correlated CFQP insulator in which the CFQPs order into a topological phase with a broken symmetry. According to existing theory, candidate ground states for this correlated CFQP insulator can be described as BPCFs, specifically the ones with one-CFQP and two-CFQPs per bubble[39,40]. There is therefore a close analogy between RIQHSs, interpreted as EBPs[13-19] and the RFQHS, interpreted as a BPCFs. In the Supplement we show that the lack of resistance anisotropy at $B = 7.76$ T is inconsistent with a stripe phase of CFQP interpretation of the RFQHS.

We now calculate the Λ-level filling factor $v^*$ at which the RFQHS forms. We first find its electronic filling factor from the electron density $n$ and the magnetic field $B$ of formation: $v = hn/eB = 1.628$. Then, $v^*$ can be obtained from the $v = 2 - v^*/(2v^* + 1)$ relation[41]. The calculated $v^* = 1.45$ value means that at the formation of the RFQHS, there is one Λ-level completely filled and another only partially filled. In this partially filled second Λ-level only a fraction $v_p^* = 0.45$ of the available states is filled. The quantity $v_p^* = 0.45$ represents the partial filling factor of the second Λ-level.

At $v^* = 1.45$, Hartree-Fock calculations predict that the BPCFs with one-CFQP per bubble and that with two-CFQPs per bubble are close in energy[39,40]. A rendering of the former is seen in Fig.1e, whereas the latter is sketched in Fig.1f. However, calculations for correlated states of high complexity, such as BPCFs, energy calculations are increasingly more difficult and drawing a conclusion on the nature of competing ground states close in energy is not straightforward. For furthering our understanding we notice that the RFQHS forms at an unusually large Λ-level filling factor. Indeed, the partial Λ-level filling factor of the RFQHS is about a factor of 4 larger than that of the WSCFs reported in Ref.(38). A larger filling factor translates into a larger CFQP density, which is conducive to a more significant overlap of the CFQP wavefunctions and thus it may lead to novel correlations that otherwise are not possible. The essence of Hartree-Fock theories is that such wavefunction overlaps afford new ways of minimizing the energy, through the formation of clusters of CFQPs, such as two-CFQP bubbles[39,40]. We think that the unusually large partial filling factor of the RFQHS favors a BPCFs with two-CFQPs per bubble interpretation.

The interpretation of the RFQHS as a BPCFs with two-CFQPs per bubble is further strengthened by a close analogy between EBPs and BPCFs. In high Landau levels there are many EBPs[18], but they are not centered to a partial filling factor significantly larger than 0.3. The only known exception for an EBPs at a large 0.44 partial filling factor is the second Landau level[16,44] and, according to Hartree-Fock calculations, these EBPs are two-electron BPs[45]. An example for such an EBP centered to a large partial filling factor is the one labeled as $R2b$ in Ref. (44). We notice that there is a close relationship between $R2b$ and the RFQHS. First, both of these BPs share their orbital quantum number of the relevant constituents. Indeed, $R2b$ forms in the second Landau level for electrons[16,44] and the RFQHS forms in the second Λ-level for CFs. Furthermore, the partial filling factor of the topmost energy level for both phases is unusually large and very close to each other: $v_p = 0.44$ for $R2b$ [44] and $v_p^* = 0.45$ for the RFQHS. The close analogy between the two phases strengthens the interpretation of the RFQHS as a BPCFs with two-CFQPs per bubble.

Ground states at filling factors related by particle-hole symmetry are typically of similar nature. It is thus worth examining the $1/3 < v < 2/5$ filling factor range, which is related to the $8/5 < v < 5/3$ range we studied by the $v \leftrightarrow 2 - v$ symmetry. Ground states in the $1/3 < v < 2/5$ range are FQHSs[46,47], such as the ones at $v = 4/11$ and $v = 3/8$. At the conjugate filling factors $v = 2 - 4/11 = 18/11$ and $v = 2 - 3/8 = 13/8$ our data does not exhibit FQHSs since the Hall resistance measured at these filling factors does not exhibit plateaus consistent with FQHSs.

Instead, we observe the BPCF at a filling factor that falls between 13/8 and 18/11. Such a drastic difference in the type of ground states at symmetry related filling factor ranges is highly unusual and it remains to be studies both theoretically and experimentally. We surmise that either a symmetry breaking term in the Hamiltonian or differences in quasiparticle interactions are likely at play. Interactions are tuned by sample parameters, such as the finite width of the quantum well and Landau level mixing and even small changes in these parameters can stabilize fundamentally different types of ground states. The transition from the paired FQHS to the stripe phase as driven by the width and Landau level mixing parameter constitutes such an example[48].

We note that theories allow for a WSCF at CFQP densities lower than that of the BPCFs with two-CFQPs per bubble. Such a phase would be signaled by full quantization, so it may already be present in our data, for example near $B = 7.65$ T in Fig.3. However, we found no transport signature that separates such a WSCFs from an Anderson insulator forming in the center of the $v = 5/3$ plateau, thus whether or not the WSCF forms near $v = 5/3$ remains an open question. A related problem in the integer quantum Hall regime is discussed in Ref.8.

It has long been recognized that the success of the CF theory in describing FQHSs hinges on the nearly independent nature of the CFQPs. Indeed, interactions between CFQPs are greatly reduced as a result of the Chern-Simmons gauge forces associated with vortex attachment countering the strong Coulomb repulsion of electrons. However, because of the discreteness of the vortex attachment process, gauge forces do not always fully compensate the Coulomb repulsion. In such instances, the residual interaction between CFs may stabilize novel ground states, such as the BPCF.

**Main References**

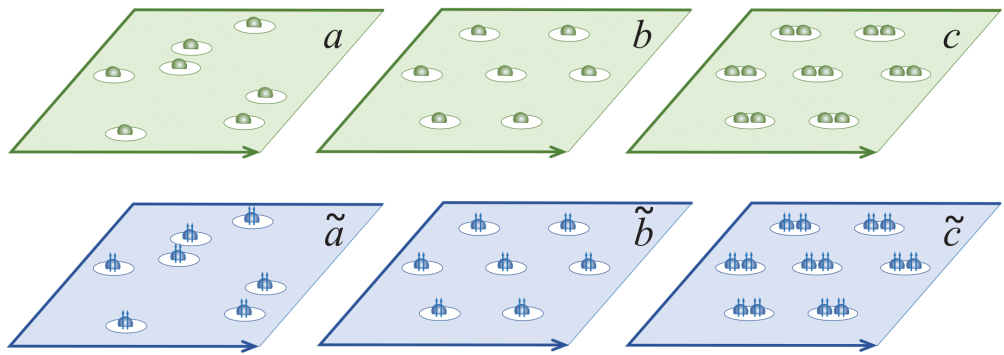

**Fig. 1. Representation of various topological phases of the electron gas with different bulk insulators.** Spheres represent quasiparticles in the valence energy level, whereas colored backgrounds represent the vacuum of the quasiparticles, i.e. completely filled lower energy levels. In the top row, electron QP based topological phases are shown. Panel $a$: an ordinary topological phase with a bulk Anderson insulator. Panels $b$ and $c$: topological phases with broken symmetry, described as electronic bubble phases with one (panel $b$) and two electrons per bubble (panel $c$), respectively. The one electron bubble phase is identical to the Wigner solid. In the bottom row, CFQP based topological phases are shown. Panel $\tilde{a}$: an ordinary topological phase with a bulk Anderson insulator. Panels $\tilde{b}$ and $\tilde{c}$: BPCFs with one- and two-CFQPs per bubble, respectively. The phase with one-CFQP per bubble is identical to the WSCFs. The two vertical arrows attached to an electron reflect the vortex attachment procedure and account for Laughlin-Jastrow correlations. Arrows along the sample boundary represent edge states.

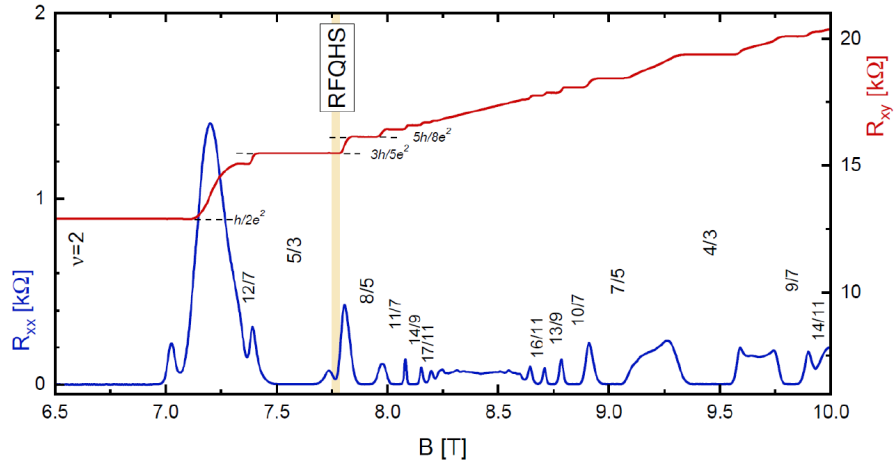

**Fig. 2. Magnetoresistance $R_{xx}$ and Hall resistance $R_{xy}$ over a broad range of magnetic fields $B$.** Traces are obtained at the temperature $T = 12\text{mK}$. Numerical labels indicate notable Landau level filling factors $\nu$ at which one integer and several fractional quantum Hall states are shown. The structure of interest associated with the reentrant fractional quantum Hall state (RFQHS) develops near $B = 7.76\text{T}$, located between Landau level filling factors $8/5 < \nu < 5/3$.

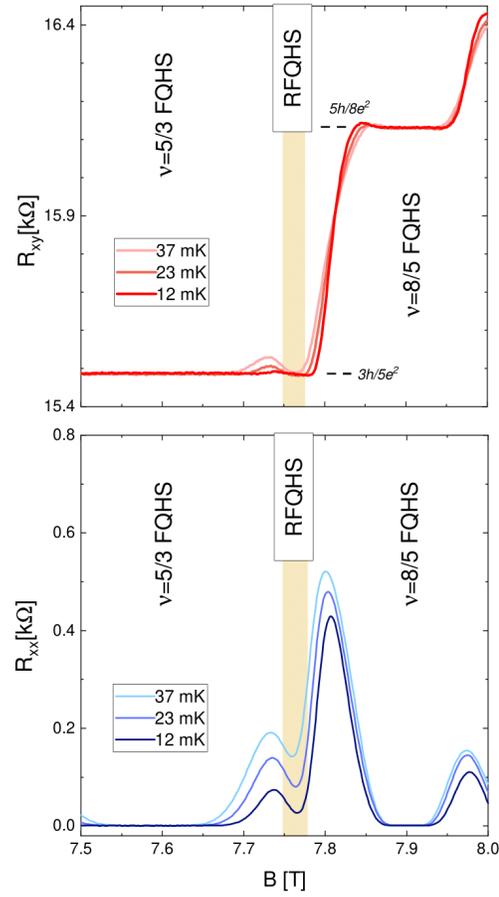

**Fig. 3. Magnetoresistance $R_{xx}$ and Hall resistance $R_{xy}$ at selected values of the temperature.** The RFQHS and the nearby $\nu = 5/3$ FQHS are separated by a conspicuous signature in magnetotrasport seen near $B = 7.73\text{T}$, both in the longitudinal and Hall resistance: a local maximum in $R_{xx}$ and a deviation from quantization in $R_{xy}$. Labels are temperatures measured in units of mK.

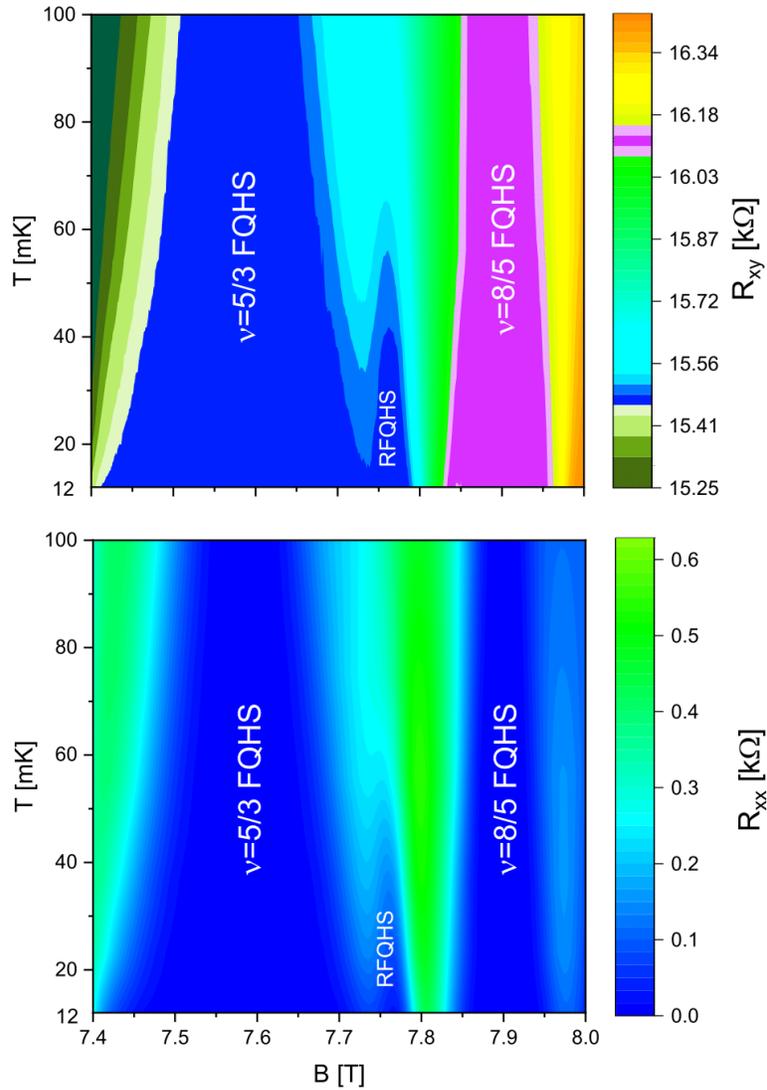

**Fig. 4. Contour plots of the magnetic field and temperature dependence of the magnetoresistance $R_{xx}$ and Hall resistance $R_{xy}$ for filling factors that include the $\nu = 5/3$ to $\nu = 8/5$ range.** The bottom panel shows $R_{xx}$, whereas the top panel $R_{xy}$. The RFQHS is observed centered to $B = 7.76$ T at temperatures as high as 60 mK.

## Methods

Data presented in the main text is generated from a sample we will refer to as Sample 1. Sample 1 is a two-dimensional electron gas confined to a 30 nm GaAs quantum well that is part of a GaAs/AlGaAs heterostructure. Doping is done in a superlattice. The density of this sample is $n = 3.06 \times 10^{11} cm^{-2}$ and the low temperature mobility is $\mu = 32 \times 10^6 cm^2/Vs$.

Magnetotransport measurements were performed in a van der Pauw sample geometry using standard lock-in technique. The excitation current used was 3 nA. Our sample was mounted in vacuum on the copper tail of our dilution refrigerator, reaching the lowest estimated temperature of $T = 12$ mK.

Sample states were prepared by illumination with a red light emitting diode located close to the sample and facing it. The illumination is performed near a sample temperature of 10 K, by passing a 4 mA current through the diode. Such an illumination slightly increases the sample density, and we believe it alleviates density inhomogeneities that may be present.

**Acknowledgments:**

We acknowledge useful discussions with Jainendra Jain and Vito Scarola. Measurements at Purdue were supported by the NSF DMR Grant No. 1904497. The sample growth effort of L.N.P., K.W.W., and K.W.B. of Princeton University was supported by the Gordon and Betty Moore Foundation EPiQS Grant No. GBMF 9615, and the National Science Foundation MRSEC Grant No. DMR 2011750.

**Author contributions:**

V.S. and A.K. performed low temperature transport measurements. L.N.P, K.W.W and K.W.B. produced MBE grown GaAs/AlGaAs samples and characterized them. V.S., H.H, and G.A.C. analyzed the data and wrote the manuscript.

**Competing interests:**

The authors declare no competing interests.

**Data and materials availability:**

Data supporting the findings of this study are available from the corresponding author upon request.

Correspondence and requests for materials should be addressed to G.A.C.